\pdfoutput=1
\documentclass[12pt]{article}
\usepackage{amsmath}
\usepackage{graphicx,psfrag,epsf}
\usepackage{enumerate}
\usepackage{natbib}
\usepackage{url} 
\usepackage{alltt,multirow,booktabs,setspace}  %
\usepackage{amssymb,amsfonts,algorithm,algpseudocode}
\newcommand{\blind}{0}

\begin{document}

\def\spacingset#1{\renewcommand{\baselinestretch}%
{#1}\small\normalsize} \spacingset{1}


\if0\blind
{
  \title{\bf Centered Isotonic Regression: Point and Interval Estimation for Dose-Response Studies}
  \author{Assaf P. Oron\hspace{.2cm}\\
    Children's Core for Biomedical Statistics, Seattle Children's Research Institute\\
    and \\
    Nancy Flournoy \\
    Department of Statistics, University of Missouri}
  \maketitle
} \fi

\if1\blind
{
  \bigskip
  \bigskip
  \bigskip
  \begin{center}

    {\LARGE\bf Centered Isotonic Regression: Point and Interval Estimation for Dose-Response Studies}
\end{center}
  \medskip
} \fi
\begin{center}
  \medskip
\emph{\large Author's version, as accepted for publication to Statistics in Biopharmaceutical Research; supplement is at end}
\end{center}

\bigskip
\begin{abstract}
Univariate isotonic regression (IR) has been used for nonparametric estimation in dose-response and dose-finding studies. One undesirable property of IR is the prevalence of piecewise-constant stretches in its estimates, whereas the dose-response function is usually assumed to be strictly increasing. We propose a simple modification to IR, called centered isotonic regression (CIR). CIR's estimates are strictly increasing in the interior of the dose range. In the absence of monotonicity violations, CIR and IR both return the original observations. Numerical examination indicates that for sample sizes typical of dose-response studies and with realistic dose-response curves, CIR provides a substantial reduction in estimation error compared with IR when monotonicity violations occur. We also develop analytical interval estimates for IR and CIR, with good coverage behavior. An R package implements these point and interval estimates.
\end{abstract}

\noindent%
{\it Keywords:}  dose-finding, nonparametric statistics, nonparametric regression, up-and-down, binary data analysis, small-sample methods
\vfill

\newpage
\spacingset{1.45} 
\section{Introduction}\label{sec:intro}

Isotonic regression (IR) is a standard constrained nonparametric estimation method \citep{BarlowEtAl72,RobertsonEtAl88}. For a comprehensive 50-year history of IR, see \citet{deLeeuwEtAl09}. The following article discusses the simplest, and probably the most common application of IR: estimating a monotone increasing univariate function $y=F(x)$, based on observations or estimates $\mathbf{y}=y_1,\ldots ,y_m$ at $m$ unique $x$ values or \emph{design points}, $x_1<\ldots <x_m$, in the absence of any known parametric form for $F(x)$. For this problem, IR is defined as a function $\hat{F}$ minimizing the sum of square errors $\sum_j \left(\hat{F}\left(x_j\right)-y_j\right)^2$, subject to the monotonicity constraint. If monotonicity is not violated, IR simply returns the original observations. In case of violation, the IR estimate includes sequences of identical values replacing the observations in violation.

We restrict ourselves here to data from dose-response studies with binary responses, with the $m$ design points fixed \emph{a priori}. In these studies, $\mathbf{y}$ represents summaries based on conditional Binomial outcomes, and IR provides the nonparametric maximum likelihood estimate (NPMLE) of $F$ under the monotonicity constraint. Dose-response studies of this type might aim to estimate $F(x)$, or to perform \emph{dose-finding}: to estimate a dose $x^*$ such that $F(x^*)=p$ for some given response rate $p$, i.e., $x^*=F^{-1}(p)$. In typical dose-response studies in medicine, pharmaceutical research, toxicology, and engineering, $m$ is small and the overall sample size $n$ is small to moderate, due to time, budget or ethical constraints.

Dose-finding requires estimates of $F$ over the continuous dose range $\left[x_1,x_m\right]$, but IR returns estimates only at the design points, with no explicit functional estimate between design points. When the goal is to estimate a cumulative distribution function (CDF), methodological convention dictates a discontinuous, piecewise-constant curve. In real-life dose-finding studies, a continuous strictly-increasing curve is more realistic and more useful. Unfortunately, under the sample-size restrictions described above, since the design points are few and fixed, and $F$'s shape unknown beyond its monotonicity, estimates of $F$ between the design points are rather speculative. Nonparametric functional estimates, e.g., constrained splines \citep{WangLi08}, which work well when $m$ and $n$ are sufficiently large, have little support to rely upon and often fail to converge.

Linear interpolation between estimates at design points is arguably the safest  solution. Note that such a solution can do no better than approximate $\overline{F}$, the linear interpolation between $F$ values at design points.  \citet{StylianouFlournoy02} suggested linear interpolation of IR, a solution that is simple and requires no additional assumptions or parameters. This generates a piecewise-linear curve, hereafter called $\widehat{\overline{F}}$. In the presence of monotonicity violations, $\widehat{\overline{F}}$ is piecewise-constant along certain stretches. This can be viewed as undesirable because it is usually assumed or even known that $F$ is strictly increasing. As we show in Section~\ref{sec:numer}, these piecewise-constant segments are also associated with poor estimation performance.

Few strictly-increasing, nonparametric or semiparametric solutions for the small-$m$ domain have been published. \citet{Schmoyer84} developed an algorithm to calculate a strictly monotone NPMLE under the restriction that $F$ is sigmoidally-shaped, and commented that sometimes the algorithm does not converge. \citet{Lee96} suggested a solution based on the average of analytically-calculated asymptotic upper and lower monotone confidence bands for $F$. \citet{IasonosOstrovnaya11} suggested to numerically find the NPMLE, under the constraint of user-provided lower and upper bounds on the increase in $F$ between adjacent design points. None of these suggestions have been widely adopted.

Another limitation of IR has been the absence of reliable interval estimates for small samples. \citet{Korn82} suggested confidence bands for IR based on Studentized maximum modulus distributions, assuming Normal errors. \citet{Schoenfeld86} and \citet{Lee96} improve upon these bands using the same general approach. Use of the bootstrap for IR intervals in dose-response experiments was explored by \citet{DilleenEtAl03} and \citet{StylianouEtAl03}. More recently, \citet{IasonosOstrovnaya11} argued that the bootstrap is inadequate for this problem, and introduced an analytical interval combining methods from \citet{Wilson27}, \citet{Morris88}, and \citet{AgrestiCoull98}. Their approach requires user-specified parameters. We are not aware of any published study of confidence intervals for IR-based estimates of $F^{-1}$.

Our article presents novel isotonic point and interval estimates for $F$ and $F^{-1}$. We offer a simple improvement to the IR point estimate, called centered isotonic regression (CIR). CIR requires no additional assumptions and no tuning parameters. At the edges of the dose range or in the absence of monotonicity violations, CIR is identical to IR. When the two differ, we demonstrate numerically that CIR provides substantially better-performing estimates of $F$ and $F^{-1}$, for small to moderate samples under most realistic situations. Our interval estimates, applicable to both CIR and IR, are analytical, using \citet{Morris88}'s formula for ordered Binomial intervals with some modifications. We also address interval estimation along the entire dose range rather than only at the design points. None of the presented interval estimates require tuning parameters. All methods are available in a new R package called \texttt{cir}.

The article's structure is as follows: the next section presents terminology and describes the pooled-adjacent-violators algorithm (PAVA) used to obtain the IR estimate. Section~\ref{sec:cir} presents and discusses the CIR algorithm. Section~\ref{sec:interval} presents and discusses interval estimates, including an accommodation for sequential designs.  Section \ref{sec:numer} numerically compares CIR and IR point-estimate performance, and evaluates interval-estimate coverage and width. The article ends with a general discussion.

\section{Basic Terminology}\label{sec:term}
\subsection{Terminology and Assumptions}

Let $\mathbf{y}=y_1,\ldots ,y_m$ be observed proportions of conditionally independent binary responses at the design points $\mathbf{x}=x_1<\ldots <x_m$, with corresponding sample sizes $\mathbf{n}=n_1,\ldots ,n_m$. In practice, $\mathbf{y}$ might signify the presence of toxicity, successful response to treatment, etc. At each design point $x_j$, $1\leq j\leq m$,

\begin{equation}\label{eq:binom}
n_jY_j \mid x_j \sim \textrm{Binomial}\left[n_j,F\left(x_j\right)\right].
\end{equation}

We assume $F$ is monotone increasing in $x$. It can be viewed as the cumulative distribution function (CDF) of response thresholds. The overall sample size is $n=\sum_j n_j$. Hereafter, the term \emph{forward} when used will refer to estimation of $F$, while \emph{inverse} and \emph{dose-finding} refer to estimation of  $F^{-1}$. We will also occasionally use $F_j, \hat{F}_j$, etc., as shorthand for $F\left(x_j\right), \hat{F}\left(x_j\right)$, etc.

\subsection{The Pooled-Adjacent-Violators Algorithm}

PAVA is the most popular and straightforward way to produce the IR estimate for the simple univariate-response case. It resolves any monotonicity violation in $\mathbf{y}$ by iteratively \emph{pooling} adjacent monotonicity-violating $y$ values and replacing them by their weighted average; hence its name.

\begin{algorithm}[!ht]
 \caption{
 Pool-Adjacent-Violators Algorithm (PAVA)}
\begin{algorithmic}\label{alg:PAVA}

\Procedure{PAVA}{$\mathbf{y},\mathbf{n}$}
\Statex
\State $\forall j=1,\ldots m, \ \ \hat{F}_j\gets y_j$
\State $\mathbb{C}\gets \emptyset$
\While {$\left(\mathbb{V}\gets\left\{j: j<m,\hat{F}_j>\hat{F}_{j+1}\right\}\right)\neq\emptyset$}
    \State $h\gets\min (\mathbb{V})$
\Statex
\Comment{$Note: \mathbb{V}$ enumerates all violations, whereas $\mathbb{C}$ is the current contiguous violation. The \textbf{if} below resets $\mathbb{C}$ when contiguity is broken.}
    \State \If{ $h\in\mathbb{C}$}  $\mathbb{C}\gets \mathbb{C}\cup\{h+1\}$
         \Else $\ \ \mathbb{C}\gets\{h,h+1\}$
    \EndIf
\Statex
\Comment{The actual assignment step is anti-climactic by comparison:}
\Statex
    \State $\forall j \in \mathbb{C},\ \  \hat{F}_j\gets \frac{\sum_{k\in\mathbb{C}}n_ky_k }{\sum_{k\in\mathbb{C}} n_k}$
\EndWhile
\Statex
\State \textbf{Return} $\mathbf{\hat{F}}=\hat{F}_1,\ldots ,\hat{F}_m$.
\EndProcedure
\end{algorithmic}
\end{algorithm}

In words, stepping up from $\left(x_1,y_1\right)$, the algorithm identifies the first monotonicity violation, and replaces the pair of violating $y$ values with two copies of their weighted average. If the next $y$ value is below this average, it replaces all 3 values by their overall weighted average, and so on until the violation's contiguity is broken. The final pooling is the minimal one needed to remove all monotonicity violations. As seen in the algorithm description, some bookkeeping is needed to ensure contiguous violations are identified and replaced correctly.

\citet{StylianouFlournoy02} introduced the piecewise-linear interpolation between the points $\left\{\left(x_1,\hat{F}_1,\right)\ldots,\left(x_m,\hat{F}_m,\right)\right\}$ for estimating $F$ over the continuous dose range $\left[x_1,x_m\right]$. We denote the resulting curve as $\widehat{\overline{F}}$. The inverse point estimate of $F^{-1}(p)$ can then be defined as the $x$ value where $\widehat{\overline{F}}$ crosses the horizontal line $y=p$ (see Figure~\ref{fig:1}).

\noindent A few notes about IR:

\begin{enumerate}
    \item If there are no monotonicity violations, the final IR estimates $\mathbf{\hat{F}}$ are identical to the raw observed frequencies $\mathbf{y}$.
    \item If violations exist, then the IR estimate \emph{must} include sequences of $x$ values over which $\hat{F}$ is constant. In other words, \textbf{in case of violation, the IR estimate lies on the boundary of the monotonicity constraint.}
     \item The design points, or any other $x$ value, play no role in the algorithm, and are therefore not part of its input. The design points' impacts are either indirect (via the ordering of the $y$'s) or after the fact (re-scaling the solution along the $x$-axis).
\end{enumerate}

Notes 2--3 indicate problems in the practical application of IR for dose-response studies. For investigators, a monotone dose-response relationship almost universally means \emph{strictly} monotone. Furthermore, with a piecewise-constant $\widehat{\overline{F}}$ the inverse estimate of $F^{-1}$ is non-unique along the flat stretches. Finally, the aspect ratio of $\widehat{\overline{F}}$ in two dimensions depends upon $\mathbf{x}$, which plays no role in deriving it.

\section{Centered Isotonic Regression (CIR)}\label{sec:cir}

\subsection{Algorithm Description}

CIR offers a technically minor, but rather useful modification of IR. For monotonicity violations involving interior design points, CIR collapses the IR estimates to a single point, whose $x$ coordinate is the sample-size-weighted average of participating design points, i.e., the exact same weighting used for the point estimates.

\begin{algorithm}[!ht]
 \caption{
 Centered Isotonic Regression (CIR) Algorithm}
\begin{algorithmic}\label{alg:cir}

\Procedure{CIR}{$\mathbf{x},\mathbf{y},\mathbf{n}$}
\Statex
\State $\forall j=1,\ldots m,\ \ \tilde{x}_j\gets x_j,\tilde{F}_j\gets y_j,\tilde{n}_j\gets n_j$
\State $\tilde{m}\gets m$
\While {$\mathbb{V}\equiv\left\{j: j<m,\left(\tilde{F}_j > \tilde{F}_{j+1}\right)
\lor \left(\tilde{F}_j=\tilde{F}_{j+1} \land \tilde{F}_j\in (0,1)\right)\right\}\neq\emptyset$}

    \State $h\gets\min (\mathbb{V})$
     \State $\tilde{F}_h\gets \frac{\tilde{n}_h\tilde{F}_h+\tilde{n}_{h+1}\tilde{F}_{h+1}}{\tilde{n}_h+\tilde{n}_{h+1}}$
    \State $\tilde{x}_h\gets \frac{\tilde{n}_hx_h+\tilde{n}_{h+1}x_{h+1}}{\tilde{n}_h+\tilde{n}_{h+1}}$
    \State $\tilde{n}_h\gets \tilde{n}_h+\tilde{n}_{h+1}$
    \State Remove point $h+1$
    \State $\tilde{m}\gets \tilde{m}-1$
\EndWhile
\Statex
\Statex
\Comment{End main loop; now exception handling if $\tilde{x}$'s range is shorter than original}
\Statex
\If {$\tilde{x}_1 > x_1$}
    \State Add point 1: $\left(x_1,\tilde{F}_1,0\right)$
    \State $\tilde{m}\gets\tilde{m}+1$
\EndIf
\If {$\tilde{x}_{\tilde{m}} < x_m$}
    \State Add point $\tilde{m}+1$: $\left(x_m,\tilde{F}_{\tilde{m}},0\right)$
    \State $\tilde{m}\gets\tilde{m}+1$
\EndIf
\Statex

\State \textbf{Return} $\mathbf{\tilde{x}},\mathbf{\tilde{F}},\mathbf{\tilde{n}}$.
\EndProcedure
\end{algorithmic}
\end{algorithm}

\textbf{The unique values of $\mathbf{\tilde{F}}$, the CIR point estimates, are always identical to the unique values of IR's $\mathbf{\hat{F}}$.} However, the $\mathbf{\tilde{F}}$ have fewer repetitions, and the associated $x$ values might differ. The algorithm's syntax is more straightforward than PAVA's, because the analogous pooling of both $x$ and $y$ values reduces the need for bookkeeping.

\begin{figure}[!ht]
\begin{center}
\includegraphics[scale=0.6,bb=0 0 1000 500,viewport=100 0 600 500]{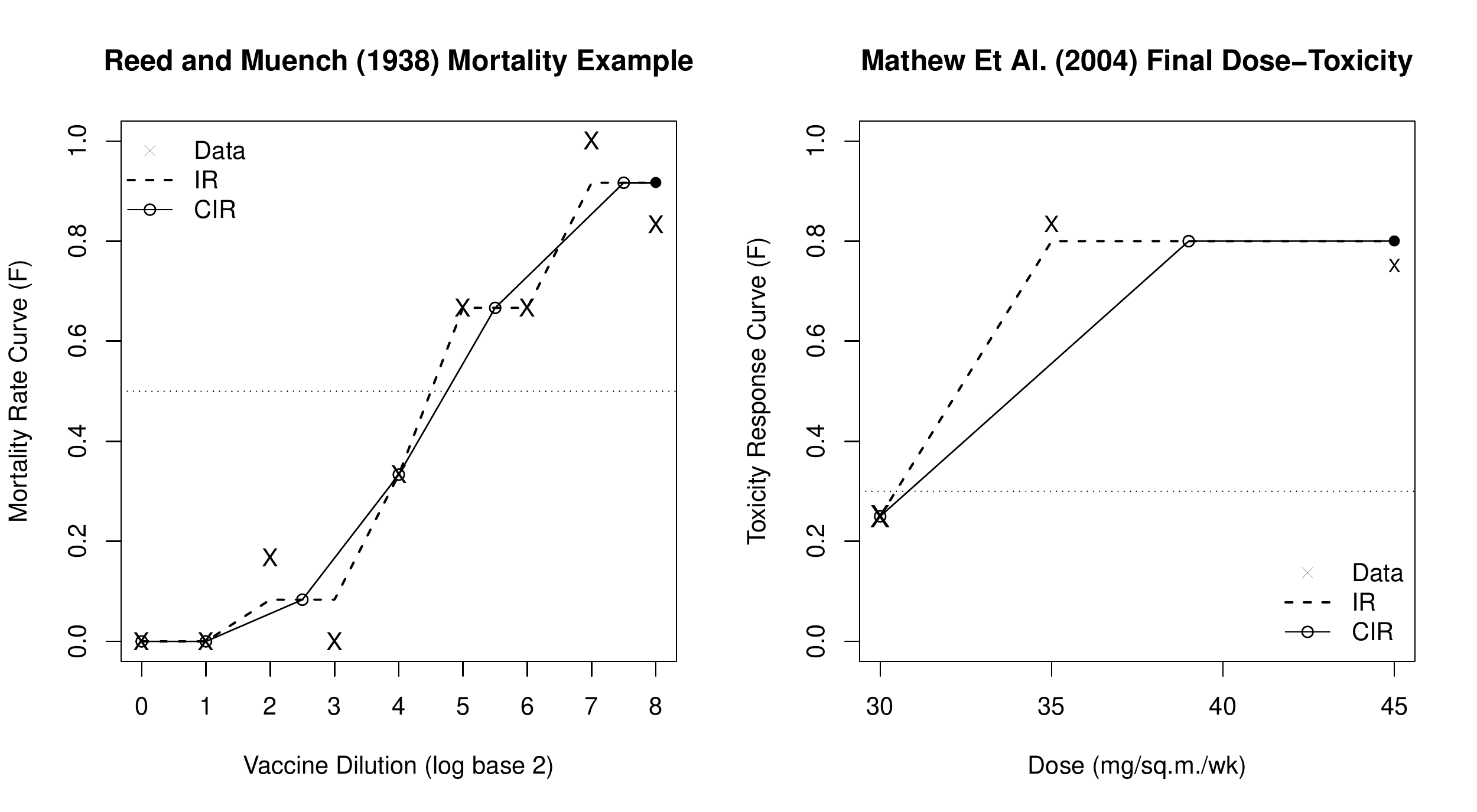}
\end{center}
\caption{Illustration of the difference between isotonic regression (IR, dashed lines) and centered isotonic regression (CIR, solid lines), using two dose-finding datasets: \citet{ReedMuench38}'s example for finding the $LD_{50}$ (an inverse estimate of the median dose), and \citet{MathewEtAl04}'s Phase~I cancer trial. The empty circles along the CIR lines indicate the points $\left(\tilde{x},\tilde{F}\right)$. The black circles indicate points added technically at the end of the CIR algorithm, in order to cover the original $x$ range. The dotted horizontal lines indicate $y=p$, i.e., each trial's target response rate. The inverse estimates for each study are the $x$ values where the IR and CIR curves cross $y=p$.}\label{fig:1}
\end{figure}

CIR can be seen as a type of shrinkage estimator, with the shrinkage occurring along the $x$-axis. Figure~\ref{fig:1} shows the raw data (`X' marks) as well as interpolated IR and CIR curves for two published datasets. When the violating stretch consists of only two points, then the location of $\left(\tilde{x}_j,\tilde{F}_j\right)$ can be found by drawing a line between the two original data points, and pinpointing its intersection with the horizontal line between IR estimates. See for example in Figure~\ref{fig:1} (left), the segment between $x=2^2$ and $x=2^3$.

CIR treats ties in $y$ values as violations, because they violate strict monotonicity. An exception is made for sequences of 0's and 1's, because these are the limits of allowed $y$ values. Note, for example, how the two 0's at the lower doses on the left-hand pane of Fig.~\ref{fig:1} are allowed to remain, while sequences of identical $y$ values at higher doses are removed. Similarly to IR, in the absence of strict monotonicity violations CIR returns the original $y$'s. Unlike PAVA, the CIR algorithm does require $x$ values as inputs.

When shrinkage along the $x$ axis takes place, the CIR algorithm might return a smaller number of points than was input. Hereafter, we denote the set of $x$ values returned by CIR as the \textbf{shrinkage points} $\mathbf{\tilde{x}}$, in contrast with the original design points. Since the CIR estimate is strictly increasing everywhere except (possibly) near the boundaries, using CIR's piecewise-linear interpolation $\widetilde{\overline{F}}$ one can obtain a unique inverse estimate of $F^{-1}$ for any value in $\left(\tilde{F}_1,\tilde{F}_{m}\right)$. The interpolation is carried out between shrinkage points rather than between the original design points.

The conditions of the \textbf{if} statements at the algorithm's end are triggered when a monotonicity violation involves a dose-range boundary. The main loop replaces the design points involved in the violation with a single shrinkage point, thereby removing $x_1$ (or $x_m$) and shortening the overall dose range. The \textbf{if} statements add flat stretches at the end as needed, extending the range back out to $[x_1,x_m]$. See the points marked with black circles in Figure~\ref{fig:1}. Along the added stretch, the CIR $\widetilde{\overline{F}}$ estimate is identical to IR's $\widehat{\overline{F}}$.

\subsection{Theoretical Motivation}\label{sec:bias}

In our view, the conceptual appeal of a strictly-monotone nonparametric estimate available at any sample size, and the practical appeal of no tuning parameters and substantially better estimation performance (as will be shown in Section~\ref{sec:numer}), suffice to make CIR a compelling alternative to IR. However, there is also some theoretical basis for CIR.

Without loss of generality, assume that monotonicity violations encompass only 2 design points labeled (for convenience) $x_1,x_2$, and the associated observations $y_1,y_2$; thanks to the algorithms' iterative nature, all the results below can be extended to larger violations via induction.

The occurrence of a monotonicity violation obviously violates a key assumption, but in practical terms this violation manifests itself as a \textbf{bias}. To see why, consider the random variable $Y_2-Y_1$. Marginally, $Y_1,Y_2$ are unbiased and $\mathrm{E}\left[Y_2-Y_1\right]>0$. Violation (in the strict, CIR sense) means $y_2\leq y_1$. The violation itself is a random event that can be conditioned upon, and thus

\begin{equation}\label{eq:bias0}
\mathrm{E}\left[Y_2-Y_1\mid Y_2\leq Y_1\right]\leq 0<\mathrm{E}\left[Y_2-Y_1\right],
\end{equation}

\noindent so there must be a \emph{conditional} bias inflicted upon either $Y_1$ or $Y_2$, or both.

\noindent Now, IR's flat stretch removes some of the bias but not all of it, because $\mathrm{E}\left[\hat{F}_2-\hat{F}_1\mid Y_2\leq Y_1\right]=0$.

Suppose we want to eliminate the conditional bias as thoroughly as possible, while remaining within the constraints of piecewise-linear interpolation. We envision shrinking violating segments onto single points, producing a curve $\widetilde{\overline{F}}$ that is strictly monotone (except possibly at the edges). What would be the optimal placement of shrinkage points along both the $x$ and $y$ axes, from a conditional-bias perspective?

Recall that, as mentioned in the Introduction, linear-interpolation estimators cannot really approximate the true nonlinear $F$ between design points.  Regardless of its placement, the theoretical bias of an interpolation estimate at a shrinkage point should be evaluated against $\overline{F}$, not $F$. This also explains why the shrinkage point's $x$ and $y$ coordinates must use the same weighting (as CIR does), since maintaining the same weights in both dimensions is what defines a linear interpolation.

Therefore, all that remains is to find an bias-eliminating weighted average of $Y_1$ and $Y_2$. Since $Y_1,Y_2$ are originally unbiased, and since violation is equivalent to $Y_2-Y_1\leq 0$, a weighted average that is uncorrelated with $Y_2-Y_1$ will retain its original expectation despite the violation. If we place it at a similarly-averaged $x$ coordinate, it will be a conditionally-unbiased estimate of $\overline{F}$ at that point. Symbolically, we are looking for $a,b$ such that

\begin{equation}\label{eq:bias1}
\mathrm{Cov}\left(aY_1+bY_2,Y_2-Y_1\right)=0;\ \ \ a,b\geq 0, a+b=1.
\end{equation}

\noindent Algebraic manipulation yields

\begin{equation}\label{eq:bias2}
\frac{b}{a}=\frac{n_2F_1\left(1-F_1\right)}{n_1F_2\left(1-F_2\right)}.
\end{equation}

In words, the bias-eliminating weights are (inverse) variance weights. If $F$ changes slowly over the violation, or if $F\approx 0.5$, then $b/a\approx n_2/n_1$, which are the PAVA weights used by IR and CIR. When this is not a good approximation, then (compared with the $n_2/n_1$ weights) more weight is given to the point whose true $F$ value is closer to the edge of the range, i.e., $F=0$ or 1. This will pull the actual estimated $\widetilde{\overline{F}}$ curve up and left towards $\left(x_1,y_1\right)$ for $F<0.5$, and vice versa for $F>0.5$, by a relatively larger amount as $F\to 0$ or 1, respectively. Conversely, this also means that CIR's $\widetilde{\overline{F}}$ using the original PAVA weights should be biased downward near $F=0$, and upwards near $F=1$.

Of course, we don't know the true $F_1$ and $F_2$ appearing in (\ref{eq:bias2}). However, a simple iterative plug-in scheme, starting with ordinary CIR to get initial $\tilde{F}$ estimates, then re-weighting observations and calling CIR again using the new weights rather than default ones, and so forth, converges fairly reliably. In our simulations, we examined this option as an alternative to the simpler $n$-weighted CIR default. In general it doesn't appear to be worth the added complication. Those results are briefly described in Section~\ref{sec:numer}.


\section{Interval Estimation}\label{sec:interval}

Interval estimation for studies of the type discussed here is rather challenging:

\begin{enumerate}
 \item Small-sample Binomial data in general have relatively poor interval estimates because of their discreteness and the mean-variance relationship \citep{BrownEtAl01}.
 \item When interpolating between design points, additional error is introduced due to $F$'s unknown curvature over the interpolation interval.
 \item For dose-finding intervals, inverse estimation requires another approximation.
 \item The most popular dose-finding designs are sequential, inducing randomness on the distribution of observations across design points, and increasing the uncertainty.
\end{enumerate}

\noindent The following discussion of confidence intervals is applicable to both IR and CIR. We use the symbol $\tilde{F}$ as reference to the point estimates from either method.

\subsection{Confidence Intervals for Forward Estimation}

Many methods constructing small-sample confidence intervals for Binomial proportions have been proposed over the years, with varying degrees of theoretical justification. \citet{BrownEtAl01} recommend any of the following three: the \citet{Wilson27} interval based on the score test, a Bayesian-motivated ``Jeffrys interval'', and a third interval they present as ``Agresti-Coull'', inspired by the ideas of \citet{AgrestiCoull98}. These analytical intervals were derived for single Binomial proportions.

\citet{Morris88} developed theoretical small-sample interval estimates for a vector of observations $\mathbf{y}$ in the presence of monotone ordering by the doses $\mathbf{x}$, and proved that interval coverage is conservative. The algorithm begins at one edge of the dose range ($x_1$ for the lower bound and vice versa), with a bound produced by inverting a hypothesis test; in the Binomial case this is known as the \citet{ClopperPearson34} bound. From $x_2$ upward, an iterative formula narrows the lower bounds using the ordering (and vice versa for the upper bounds from $x_{m-1}$ downward). There is a specific algorithm for Binomial responses, as well as a generic algorithm which requires tuning parameters. \citet{IasonosOstrovnaya11} adapted the generic algorithm, while we utilize the specific one. Since the starting bounds to Morris' algorithm are conservative, and the amount of narrowing varies by design point, there is potential room for further narrowing. Therefore, our implementation optionally allows for pointwise replacement of the Morris bounds with one of the three above-mentioned intervals. We prefer the \citet{Wilson27} interval, having the strongest theoretical justification. Further details are provided in the Supplement.

We note that IR and CIR are in violation of one of Morris' assumptions, because due to the pooling their point estimates are not conditionally independent. In general the dependence should lead to increased precision and smaller standard errors. Therefore, Binomial-based analytical intervals for IR/CIR should be somewhat conservative.

For forward intervals at values of $x$ between design points, we linearly interpolate interval boundaries between confidence bounds calculated at the design points. This approach might be anti-conservative, because $F$ is generally assumed to be nonlinear. However, as shall be shown in Section~\ref{sec:numer}, the effect as far as it exists, is dwarfed by the initial conservatism of confidence bounds at design points.

\subsection{Intervals for Inverse Estimates}

\begin{figure}[!ht]
\begin{center}
\includegraphics[scale=0.6,bb=0 0 1000 500,viewport=100 0 600 500]{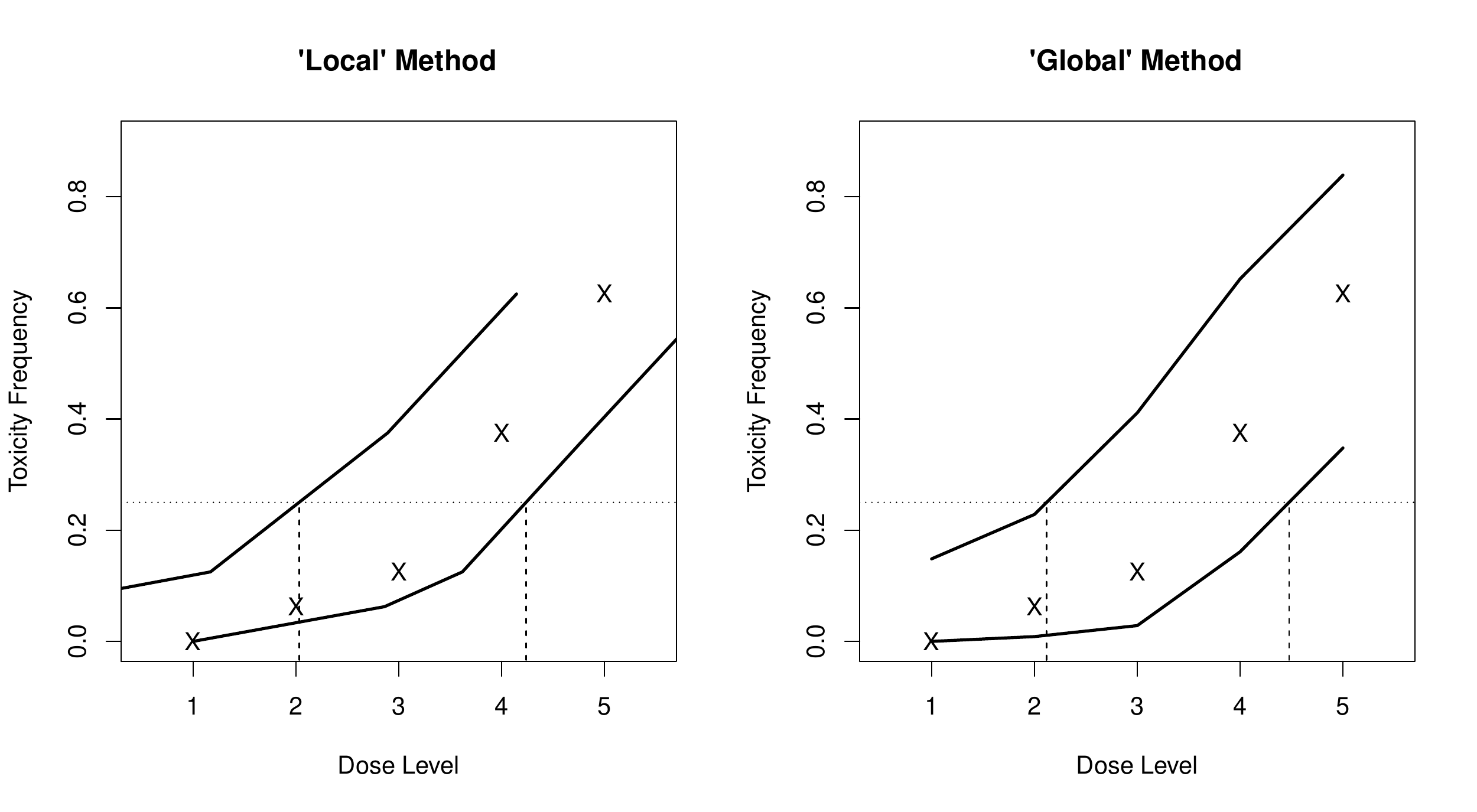}
\end{center}
\caption{Illustration of the ``local'' (left) and ``global'' (right) approaches to construct inverse confidence intervals for dose-finding. Shown are CIs for the 25th percentile. The ``local'' approach `rotates' the forward bounds at each observation, using a Delta-method based formula. Therefore, CIs are available for percentiles between the smallest and largest $\tilde{F}$, i.e., for all percentiles for which a point estimate was found. In the right-hand pane, the solid lines mark the forward confidence bounds. The ``global'' approach simply records where they cross $y=0.25$. Finite ``global'' intervals are available only for percentiles at which the two forward bounds overlap horizontally. }\label{fig:inv}
\end{figure}

For inverse interval estimation, one can begin with the forward intervals, assume that despite the small $n$ their widths are roughly proportional to the square-root of the variance, and apply basic calculus to estimate the variance of $F^{-1}$. Specifically, we use the Delta method based approximation $Var\left[g\left(F(x)\right)\right]\approx \left(g^{'}\left(F(x)\right)\right)^2$ for  $g\left(F(x)\right)=F^{-1}(p)$, and the chain-rule formula for the derivative of the inverse, $\left(F^{-1}\right)^{'}=1/F^{'}$, to conclude that a reasonable inverse interval may be obtained by dividing the width of the forward interval at each $x$ value, by an estimate of $F$'s local slope (obtained via $\widetilde{\overline{F}}$).\footnote{At design points we take the average of slopes from the two adjoining segments.} Hereafter we call this the ``local'' approach to inverse intervals (Figure~\ref{fig:inv}, left). One practical concern, especially with IR, is that the local slope might be zero. When this occurs, we expand the frame of reference beyond the local segment of $\widetilde{\overline{F}}$, and calculate an average slope over a larger range of $x$ values instead. This procedure yields a positive slope estimate, except when all forward point estimates are identical.

A simpler approach uses forward intervals directly, drawing horizontal lines between them to find the inverse interval. Thus, the interval for $F^{-1}(p)$ is defined by the intersection between the forward intervals and the $y=p$ line (Figure~\ref{fig:inv}, right). This is similar to the manner in which CIs for percentiles of survival curves are often calculated \citep[e.g.,][]{FayEtAl13}. This method, which we call ``global'', draws its validity from the monotonicity of $F$, and is generally more conservative than the ``local'' interval. It also can fail to yield a finite interval if the upper or lower forward bound do not cross $y=p$. Performance of the two approaches is compared in Section~\ref{sec:siminterval}.

\subsection{Interval Estimation with Stochastic Dose Allocation}

As explained in Sections~\ref{sec:intro}-\ref{sec:term}, we assume that the design points $\mathbf{x}$ are fixed, and the tallied responses $\mathbf{y}$ are Binomial conditional upon doses. However, the manner of \emph{assigning} doses  to individual binary observations may be stochastic. In fact, many dose-response studies and dose-finding studies use randomized, adaptive or sequential designs. In studies of this type, the conditional-Binomial variance of $Y$ is augmented by the variability of the dose-allocation process, which makes the sample sizes $\mathbf{n}$ random. A generic expression for the overall variance at any design point $x_j$, is

\begin{equation}\label{eq:seqvar1}
Var\left(Y_j\right)=Var\left(E\left[Y_j\mid n_j\right]\right) + E\left[Var\left(Y_j\mid n_j\right)\right]
\end{equation}
$$ = 0 + E\left[\frac{F_j(1-F_j)}{n_j}\right].$$

Since $1/n_j$ is convex, we are assured by Jensen's inequality that the stochastic-design variance is larger than the fixed-design variance. A rough first-order estimate for the amount of variance inflation can be found by approximating the sequential process as a purely random draw, in which the probability of assigning dose $d_j$ is some constant $\pi_j\in (0,1)$. Then by Taylor expansion,

\begin{equation}\label{eq:seqvar2}
E\left[\frac{1}{n_j}\right]\approx\frac{1}{n\pi_j} + \frac{1}{2}\left(\frac{1}{n_j}\right)^{''} \bigg|_{n_j=n\pi_j}E\left[\left(n_j-E\left[n_j\right]\right)^2\right]
\end{equation}
$$=\frac{1}{n\pi_j} + \frac{Var(n_j)}{n^3\pi_j^3}= \frac{1}{n\pi_j}\left(1+\frac{1-\pi_j}{n\pi_j}\right).$$
A straightforward plug-in estimate for $\pi_j$ is the observed allocation proportion $n_j/n$. Using this estimate, the final expression simplifies to the fixed-design variance (corresponding to the `1' in the parentheses) plus a correction term. As $\pi_j\to 1$ (deterministic allocation to the observed dose only), the correction term tends to zero. As $\pi_j\to 0$ it tends to infinity, but with $n_j=0$ there are no observations to correct for anyway.

\section{Numerical Examination}\label{sec:numer}
\subsection{Methodology}

\begin{figure}[!ht]
\begin{center}
\includegraphics[scale=0.6,bb=0 0 1000 500,viewport=100 0 600 500]{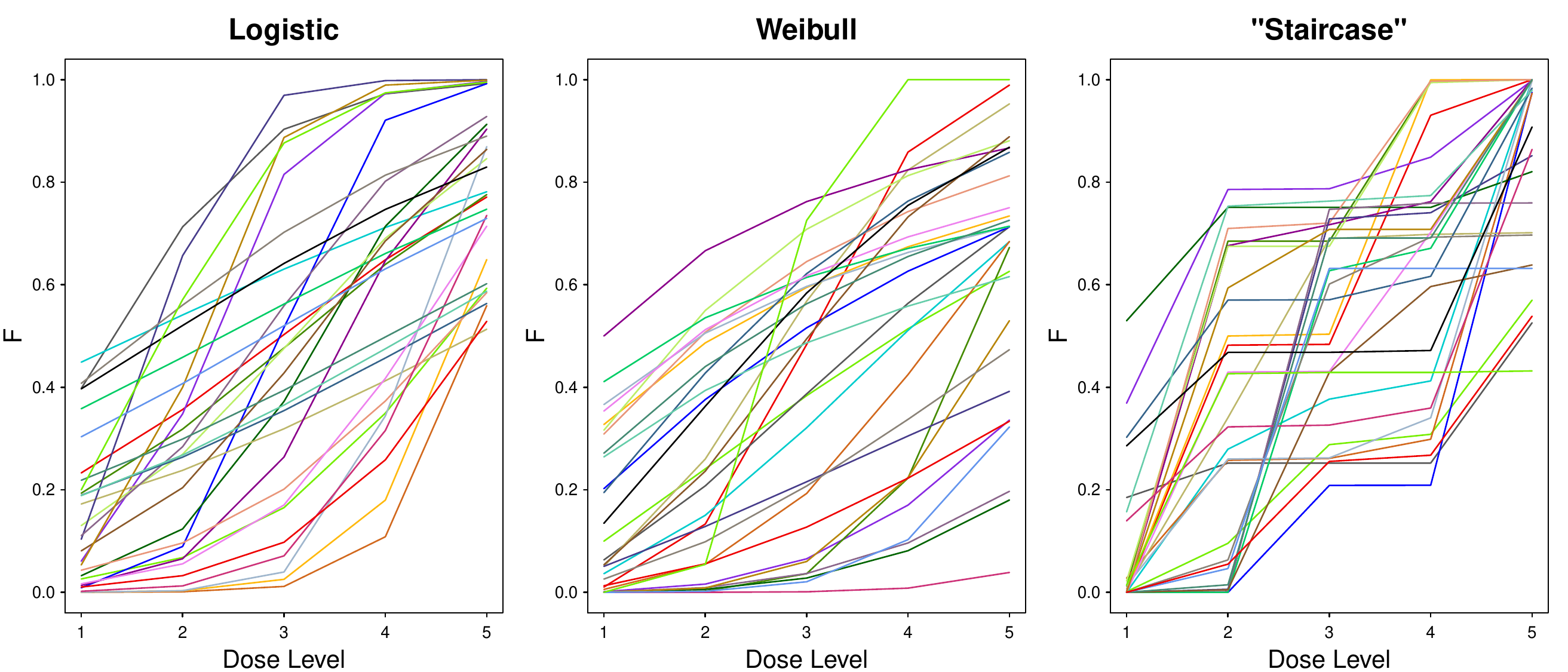}
\end{center}
\caption{Thirty randomly-selected scenarios out of the $3000$ used for the forward simulation, for each of the scenario families. Scenarios are generated by randomly drawing distribution parameters, within constraints that give preference to realistic and non-degenerate curves. Each scenario is used for a single numerical run, and performance is averaged across all scenarios in each family.}\label{fig:scen}
\end{figure}

Our simulations assume that $F$ is a CDF, and examine three families of curves:  Logistic, Weibull, and a mixture of two Normals parameterized and vetted so that $F$ will have a staircase-like shape. The former two parameterizations are commonly encountered in dose-response studies. We added the third family, hereafter called ``Staircase'', in order to generate `pathological', rarely-encountered curves whose shape is similar to the piecewise-constant IR output, and for which CIR's shrinkage of the constant stretches might be detrimental. Within each family, an ensemble of many curves (also called ``scenarios'') is generated by drawing the distribution parameters randomly. Each curve is used for a single random ``experiment'' (also called a ``run''), and overall statistics are calculated across the entire ensemble of runs. This simulation approach has become more popular recently \citep{PaolettiEtAl04,Azriel12,OronHoff13}. It presents a more realistic glimpse of the variability observed in practice, compared to the traditional simulation approach in which ensembles of random experiments are drawn from a few consciously-selected curves. Figure~\ref{fig:scen} displays subsets of the random curves used; note that it shows $\overline{F}$, rather than the underlying nonlinear $F$. All simulations had $m=5$ evenly-spaced design points, and sample sizes of $n=20,40$ and $80$.

For point estimation (Tables \ref{tbl:fwd}--\ref{tbl:seq}), the right columns of each table show the empirical root-mean-square error (RMSE) at selected $x$ values and quantiles, for IR and CIR. The left columns provide summary statistics averaged across all values from the right-hand side: the percent of estimates for which the IR and CIR estimates differed, and the average ratio between mean-square-errors (MSEs) of IR and CIR, calculated only using runs in which the IR and CIR point estimates differed. For example, an average MSE ratio of 2 indicates that on average, across all values tabulated on the right-hand side and all runs in which IR and CIR differed, the MSEs of CIR point estimates were half as large as those of IR estimates (in other words, the RMSEs differed by a factor of $\sqrt{1/2}=0.707$). With fixed design points, CIR and IR converge at a $\sqrt{n}$ rate. Therefore the MSE ratio roughly indicates the equivalent sample-size savings when switching from IR to CIR, if a monotonicity violation had occurred.

For interval estimates  (Tables \ref{tbl:fwdcov1}--\ref{tbl:seqcov}) only CIR's statistics are shown. The main statistics are average coverage rate and interval width (the latter presented in parentheses). We examined $90\%$ CIs rather than the more commonly used $95\%$, because for small-sample binary data $95\%$ is often too ambitious, and also because $90\%$ better conveys the degree of interval conservatism in case of over-coverage, which, as seen below, is rather common with the methods examined. Additional details for each specific simulation are provided when relevant.

\subsection{Performance of Point Estimates}
\subsubsection{Forward Point Estimation}

In this simulation the sample was equally split between the design points. For each family, $N=3000$ curves were generated. On the boundaries of the dose range, CIR and IR point estimates are always identical. Therefore, Table~\ref{tbl:fwd} only summarizes point estimates at the three interior design points, as well as at two intermediate points chosen in order to examine interpolation performance: one halfway between $x_2$ and $x_3$ (``$x_{2.5}$''), and the other three-quarters of the way between $x_3$ and $x_4$ (``$x_{3.75}$'').

For the two standard curve families (Logistic and Weibull), CIR performs substantially better, with MSE ratios around 2. CIR's advantage increases somewhat as $n$ increases. However, as expected, the proportion of runs with differing IR and CIR estimates decreases with increasing sample size. Comparing the estimates at design points with the interpolated estimates (two rightmost columns), with CIR the errors are about the same, whereas with IR the interpolation errors are substantially smaller than errors at design points, albeit still greater than CIR errors.
The Staircase family exhibits a diametrically opposite pattern.

Analogous summaries of empirical bias are available in the Supplement. As explained in Section~\ref{sec:bias}, when $F$ is close to 0 we expect CIR to be negatively biased, and vice versa. This is borne out for the Logistic and Weibull scenarios; despite scenarios having different $F$ curves, on average $x_2<0.5$ and $x_4>0.5$ most of the time, and indeed the bias is in the expected direction. Bias decreases as $n$ increases, perhaps because the raw Binomial $y$ values become more fine-grained. IR bias usually has the opposite sign. For Logistic and Weibull, even the worst biases are an order of magnitude smaller than overall RMSE, demonstrating that variance dominates estimation error. For the ``staircase'' family, CIR bias is roughly equal to variance, even greater as $n$ increases. This is not due to incorrect weighting, but rather because this family was chosen intentionally to be incompatible with CIR's shape constraints.

In Section \ref{sec:bias}, we described improving CIR's bias-correction weights via an iterative scheme. We employed this scheme for forward estimation (summaries not shown). For Logistic and Weibull scenarios, bias was reduced by as much as one-third, but overall RMSEs decreased only by $\sim 2\%$, at most; often, in particular near $x_3$, RMSEs even increased. For the ``staircase'' family the iterative scheme made no difference. Keeping in mind that variance dominates bias, and that the ``plug-in'' estimates used in this scheme are themselves imprecise and therefore the increase in variance might offset some of the modest bias improvements, we concluded that sticking with the fixed PAVA sample-size weights is more preferable overall.

\begin{table}[!ht]
\centering
\begin{tabular}{cccrrrrrrr}
  \toprule
   \multicolumn{3}{c}{Conditions} & \multicolumn{2}{c}{Overall Statistics} & \multicolumn{5}{c}{Pointwise RMSEs at $x$ Values} \\
 Family & $n$ & Method & \%Unequal & \textbf{MSE Ratio} & $x_2$  & $x_3$ & $x_4$ & ``$x_{2.5}$'' & ``$x_{3.75}$'' \\
  \midrule
\parbox[t]{0mm}{\multirow{6}{*}{\rotatebox[origin=c]{90}{Logistic}}} & \multirow{2}{*}{20} & IR & \multirow{2}{*}{49.0} & \multirow{2}{*}{\textbf{1.92}} & 0.18 & 0.17 & 0.18 & 0.14 & 0.14 \\
& & CIR & & & 0.12 & 0.11 & 0.12 & 0.12 & 0.12 \\
   & \multirow{2}{*}{40} & IR & \multirow{2}{*}{46.6} & \multirow{2}{*}{\textbf{1.98}} & 0.13 & 0.13 & 0.14 & 0.10 & 0.11 \\
 & & CIR & &  & 0.09 & 0.08 & 0.09 & 0.09 & 0.09 \\
   & \multirow{2}{*}{80} & IR & \multirow{2}{*}{36.5} & \multirow{2}{*}{\textbf{2.31}} & 0.10 & 0.11 & 0.10 & 0.08 & 0.08 \\
 & & CIR  &  &  & 0.06 & 0.06 & 0.06 & 0.07 & 0.07 \\
  \midrule
\parbox[t]{0mm}{\multirow{6}{*}{\rotatebox[origin=c]{90}{Weibull}}} & \multirow{2}{*}{20} & IR & \multirow{2}{*}{46.6} & \multirow{2}{*}{\textbf{1.77}}  & 0.18 & 0.17 & 0.17 & 0.13 & 0.14 \\
 & & CIR  &  &  & 0.13 & 0.11 & 0.12 & 0.12 & 0.12 \\
   & \multirow{2}{*}{40} & IR & \multirow{2}{*}{45.7} & \multirow{2}{*}{\textbf{2.03}} & 0.13 & 0.13 & 0.13 & 0.10 & 0.10 \\
 & & CIR  &  &  & 0.09 & 0.08 & 0.08 & 0.08 & 0.09 \\
  & \multirow{2}{*}{80} & IR & \multirow{2}{*}{38.7} & \multirow{2}{*}{\textbf{2.13}} & 0.11 & 0.10 & 0.09 & 0.07 & 0.08 \\
 & & CIR  &  &  & 0.07 & 0.06 & 0.06 & 0.06 & 0.06 \\
     \midrule
\parbox[t]{0mm}{\multirow{6}{*}{\rotatebox[origin=c]{90}{``Staircase''}}} & \multirow{2}{*}{20} & IR & \multirow{2}{*}{57.8} & \multirow{2}{*}{\textbf{0.76}} & 0.17 & 0.16 & 0.16 & 0.15 & 0.15 \\
  & & CIR &  &  & 0.20 & 0.18 & 0.19 & 0.17 & 0.17 \\
  & \multirow{2}{*}{40} & IR & \multirow{2}{*}{57.9} & \multirow{2}{*}{\textbf{0.60}}& 0.12 & 0.12 & 0.12 & 0.13 & 0.12 \\
  & & CIR &  &  & 0.18 & 0.16 & 0.17 & 0.15 & 0.14 \\
  & \multirow{2}{*}{80} & IR & \multirow{2}{*}{52.9} & \multirow{2}{*}{\textbf{0.45}} & 0.09 & 0.09 & 0.09 & 0.11 & 0.09 \\
  & & CIR &  &  & 0.17 & 0.14 & 0.15 & 0.14 & 0.12 \\
   \bottomrule
\end{tabular}
\caption{Forward point-estimation summaries, arranged by curve family and $n$.}\label{tbl:fwd}
\end{table}

\subsubsection{Dose-Finding, Fixed Design}

Here, too, the sample was split equally among design points. Table~\ref{tbl:bwd} presents peformance for estimates of the $25$th and $50$th percentiles of $F$. Since the proportion of runs in which IR and CIR differed was smaller, ensemble size was increased to $N=5000$.

For the Logistic and Weibull families, patterns are similar to those of Table~\ref{tbl:fwd}, albeit less dramatic. Interestingly, here CIR performs better than IR for the Staircase family curves as well, except at $n=80$.

\begin{table}[!ht]
\centering
\begin{tabular}{cccrrrr}
  \toprule
  \multicolumn{3}{c}{Conditions} & \multicolumn{2}{c}{Overall Statistics} & \multicolumn{2}{c}{Pointwise RMSEs} \\
 Family & $n$ & Method & \%Unequal & \textbf{MSE Ratio} & $F^{-1}(0.25)$ & $F^{-1}(0.5)$ \\
  \midrule
\parbox[t]{0mm}{\multirow{6}{*}{\rotatebox[origin=c]{90}{Logistic}}} & \multirow{2}{*}{20} & IR & \multirow{2}{*}{45.2} & \multirow{2}{*}{\textbf{1.66}} & 0.81 & 0.68 \\
  & & CIR  &  &  & 0.58 & 0.58 \\
  & \multirow{2}{*}{40} & IR & \multirow{2}{*}{41.6} & \multirow{2}{*}{\textbf{1.73}} & 0.67 & 0.56 \\
  & & CIR  &  &  & 0.47 & 0.47 \\
  & \multirow{2}{*}{80} & IR & \multirow{2}{*}{30.2} & \multirow{2}{*}{\textbf{1.67}} & 0.55 & 0.47  \\
  & & CIR  &  &  & 0.40 & 0.39 \\
   \midrule
\parbox[t]{0mm}{\multirow{6}{*}{\rotatebox[origin=c]{90}{Weibull}}} & \multirow{2}{*}{20} & IR & \multirow{2}{*}{33.4} & \multirow{2}{*}{\textbf{1.87}} & 0.81 & 0.70 \\
  & & CIR  &  &  & 0.58 & 0.52 \\
  & \multirow{2}{*}{40} & IR & \multirow{2}{*}{27.4} & \multirow{2}{*}{\textbf{1.96}} & 0.67 & 0.59 \\
  & & CIR &  &  & 0.48 & 0.42 \\
  & \multirow{2}{*}{80} & IR & \multirow{2}{*}{18.3} & \multirow{2}{*}{\textbf{1.94}} & 0.51 & 0.55 \\
  & & CIR &  &  & 0.37 & 0.39 \\
     \midrule
\parbox[t]{0mm}{\multirow{6}{*}{\rotatebox[origin=c]{90}{``Staircase''}}} & \multirow{2}{*}{20} & IR & \multirow{2}{*}{59.5} & \multirow{2}{*}{\textbf{1.32}} & 0.96 & 0.96  \\
  & & CIR &  &  & 0.77 & 0.92  \\
  & \multirow{2}{*}{40} & IR & \multirow{2}{*}{58.6} & \multirow{2}{*}{\textbf{1.11}} & 0.77 & 0.83 \\
  & & CIR &  &  & 0.69 & 0.84 \\
  & \multirow{2}{*}{80} & IR & \multirow{2}{*}{55.2} & \multirow{2}{*}{\textbf{0.94}} & 0.66 & 0.67 \\
  & & CIR &  &  & 0.63 & 0.76 \\
   \bottomrule
\end{tabular}
\caption{Inverse point-estimation summaries, by curve family and $n$.}\label{tbl:bwd}
\end{table}

\subsubsection{Dose-Finding, Sequential Design}

We use an up-and-down design, the family of methods for which \citet{StylianouFlournoy02} introduced IR interpolation as an estimator. The particular design in the simulations is known as `Geometric' \citep{Gezmu96} or `$k$-in-a-row' \citep{IvanovaEtAl03,OronHoff09}, with $k=2$. The design escalates by one design point after 2 consecutive negative responses at the current dose, de-escalates after any positive response, and remains at the same dose after the first negative response. Asymptotically, this dose-allocation process converges to a random walk centered slightly below $F^{-1}(0.29)$, adequate for estimating the 25th to 33th percentiles. Table~\ref{tbl:seq} presents performance statistics for estimates of the 30th percentile. Ensemble size was $N=5000$. Patterns are similar to those of Table~\ref{tbl:bwd}, but with the differences between CIR and IR less pronounced.

\begin{table}[!ht]
\centering
\begin{tabular}{ccrr}
  \toprule
 Family & $n$ &  \%Unequal & \textbf{MSE Ratio} \\
  \midrule
\multirow{3}{*}{Logistic} & 20 &  28.0 & \textbf{1.46} \\
  & 40 &  31.1 & \textbf{1.65} \\
  & 80 & 25.0 & \textbf{1.64} \\
  \midrule
\multirow{3}{*}{Weibull} & 20 &  24.8 & \textbf{1.47} \\
  & 40 &  26.4 & \textbf{1.51} \\
  & 80 & 22.8 & \textbf{1.50} \\
\midrule
  \multirow{3}{*}{``Staircase''} & 20 & 22.8 & \textbf{1.23} \\
  & 40 & 26.2 & \textbf{1.15} \\
  & 80 & 27.6 & \textbf{1.06} \\
\bottomrule
\end{tabular}
\caption{Sequential dose-finding point-estimate summaries, arranged by curve family and $n$.}\label{tbl:seq}
\end{table}

\subsection{Confidence-Interval Coverage}\label{sec:siminterval}

We evaluate interval estimates at design points separately from estimates at interpolation points, due to the possibility of poorer coverage at the latter, where observations are not available.

\subsubsection{Forward Intervals at design points}

Table~\ref{tbl:fwdcov1} compares our recommended approach, which uses the \citet{Morris88} bounds with additional narrowing using the \citet{Wilson27} bounds, to each of its components alone.  Summary statistics are averaged across all 5 design points. All methods are rather conservative, but our combined approach is the least conservative, and also has the narrowest intervals. The conservatism decreases somewhat with increasing sample size, and is also less pronounced for the Staircase scenarios.

\begin{table}[!ht]
\centering
\begin{tabular}{ccrrr}
  \toprule
 Family & $n$ &  Combined Interval & Morris & Wilson \\
  \midrule
\multirow{3}{*}{Logistic} & 20 & 0.97 (0.47) & 0.99 (0.52) & 0.98 (0.52) \\
 & 40 & 0.96 (0.37) & 0.99 (0.40) & 0.97 (0.40) \\
 & 80 & 0.95 (0.28) & 0.98 (0.30) & 0.95 (0.29) \\
\midrule
\multirow{3}{*}{Weibull} & 20 & 0.97 (0.46) & 0.99 (0.50) & 0.98 (0.51) \\
 & 40 & 0.97 (0.36) & 0.99 (0.39) & 0.97 (0.39) \\
 & 80 & 0.96 (0.27) & 0.98 (0.29) & 0.96 (0.28) \\
\midrule
  \multirow{3}{*}{``Staircase''} & 20 & 0.96 (0.47) & 0.98 (0.52) & 0.97 (0.51) \\
  & 40 & 0.93 (0.36) & 0.96 (0.4) & 0.94 (0.38) \\
  & 80 & 0.91 (0.26) & 0.93 (0.29) & 0.91 (0.27) \\
\bottomrule
\end{tabular}
\caption{Forward interval coverage summaries for $90\%$ CIs at design points, for the same simulations used in point estimation (Table~\ref{tbl:fwd}), arranged by curve family and $n$. In parentheses, the intervals' average width.}\label{tbl:fwdcov1}
\end{table}

\subsubsection{Forward Intervals at Interpolation Points}

Table~\ref{tbl:fwdcov2} compares the same 3 methods of Table~\ref{tbl:fwdcov1}, but at the two interpolation points used for Table~\ref{tbl:fwd}. The intervals are almost as conservative as at design points, except with the Staircase family, for which intervals often lack in coverage, a behavior that will be encountered again below.

\begin{table}[!ht]
\centering
\begin{tabular}{ccrrr}
  \toprule
 Family & $n$ &  Combined Interval & Morris & Wilson \\
  \midrule
\multirow{3}{*}{Logistic} & 20  & 0.97 (0.51) & 0.99 (0.56) & 0.98 (0.54) \\
  & 40 &   0.96 (0.40) & 0.98 (0.43) & 0.96 (0.42) \\
  & 80 &  0.95 (0.30) & 0.97 (0.33) & 0.95 (0.31) \\
\midrule
\multirow{3}{*}{Weibull} & 20 &  0.97 (0.49) & 0.98 (0.54) & 0.97 (0.53) \\
  & 40 &   0.96 (0.38) & 0.98 (0.41) & 0.96 (0.41) \\
  & 80 &  0.94 (0.29) & 0.96 (0.31) & 0.94 (0.30) \\
\midrule
  \multirow{3}{*}{``Staircase''} & 20 &  0.92 (0.53) & 0.95 (0.60) & 0.93 (0.55) \\
  & 40 &  0.87 (0.42) & 0.9 (0.46) & 0.87 (0.43) \\
  & 80 &  0.80 (0.32) & 0.83 (0.34) & 0.81 (0.32) \\
\bottomrule
\end{tabular}
\caption{Forward interval coverage summaries for $90\%$ CIs at interpolation points.}\label{tbl:fwdcov2}
\end{table}

\subsubsection{Inverse Intervals, Fixed Design}

For inverse interval estimation performance, we present results only from the intervals produced via the combined approach, because in Tables~\ref{tbl:fwdcov1}--\ref{tbl:fwdcov2} it consistently exhibited the narrowest intervals with coverage closest to $90\%$. Table~\ref{tbl:bwdcov} compares the ``local'' and ``global'' approaches for inverting these intervals (see Fig.~\ref{fig:inv}).

The second and third columns from the left summarize how often an inverse interval can be calculated at all, while avoiding extrapolation outside observed data. The ``local'' approach does far better than the ``global'' approach on this metric. The small minority of runs for which a ``local'' interval is not found, corresponds to cases in which the point estimate does not exist, or in which all point estimates are identical and therefore the slope is zero everywhere.

Both approaches exhibit conservative coverage, but the ``local'' approach is closer to the nominal $90\%$. Under the Staircase scenarios coverage is deficient. The width calculations only include runs in which both intervals were available. The ``local'' intervals were substantially narrower.

\begin{table}[!ht]
\centering
\begin{tabular}{ccrrrr}
  \toprule
  & & \multicolumn{2}{c}{Proportion Found} & \multicolumn{2}{c}{Coverage (width)} \\
 Family & $n$ &  ``Local'' & ``Global'' & ``Local'' & ``Global''  \\
  \midrule
\multirow{3}{*}{Logistic} & 20 & 95\%  & 45\% & 0.93 (2.18) & 0.98 (3.10)  \\
  & 40 & 96\%  & 70\% & 0.95 (1.99) & 0.98 (2.38) \\
  & 80 & 97\%  & 86\% & 0.95 (1.67) & 0.97 (1.87) \\
\midrule
\multirow{3}{*}{Weibull} & 20 & 97\%  & 55\% & 0.92 (1.96) & 0.97 (2.58)  \\
  & 40 & 97\%  & 79\% & 0.93 (1.62) & 0.96 (1.81) \\
  & 80 & 98\%  & 92\% & 0.93 (1.33) & 0.96 (1.42) \\
\midrule
  \multirow{3}{*}{``Staircase''} & 20 & 99\%  & 64\% & 0.84 (2.03) & 0.93 (2.82)  \\
  & 40 & 99\%  & 90\% & 0.81 (1.67) & 0.83 (1.93) \\
  & 80 & 99\%  & 97\% & 0.74 (1.34) & 0.69 (1.43) \\
\bottomrule
\end{tabular}
\caption{Performance of $90\%$ inverse (dose-finding) CIs, for the same simulations used to study point estimation, arranged by curve family and $n$.}\label{tbl:bwdcov}
\end{table}

\subsubsection{Inverse Interval Estimation, Sequential Design}

Given the decisive results in favor of the ``local'' interval estimates in Table~\ref{tbl:bwdcov}, we only present results for ``local'' intervals in Table~\ref{tbl:seqcov} which shows performance with a sequential dose-finding design.  Here we examined how much coverage is lost due to the random allocation, and whether the first-order correction (\ref{eq:seqvar2}) helps sustain coverage. At larger sample sizes, coverage is close to the nominal $90\%$, but it is deficient for $n=20$. The uncertainty correction only adds about $1\%$. As before, coverage on the Staircase scenarios is substantially lower.

\begin{table}[!ht]
\centering
\begin{tabular}{ccrr}
  \toprule
 Family & $n$ &  ``Local'' & ``Local''+Sequential   \\
  \midrule
\multirow{3}{*}{Logistic} & 20 & 0.80 (2.14) & 0.81 (2.25)  \\
  & 40 & 0.88 (2.02) & 0.88 (2.07)  \\
  & 80 & 0.92 (1.63) & 0.93 (1.65)  \\
\midrule
\multirow{3}{*}{Weibull} & 20 & 0.80 (1.93) & 0.81 (2.02)  \\
  & 40 & 0.88 (1.85) & 0.88 (1.90) \\
  & 80 & 0.91 (1.58) & 0.91 (1.60)  \\
\midrule
  \multirow{3}{*}{``Staircase''} & 20 & 0.75 (1.30) & 0.77 (1.35)  \\
  & 40 & 0.77 (1.11) & 0.77 (1.13)  \\
  & 80 & 0.75 (0.88) & 0.75 (0.89) \\
\bottomrule
\end{tabular}
\caption{Performance of $90\%$ ``local'' inverse interval for the up-and-down design used for point estimation (Table~\ref{tbl:seq}), with and without the sequential-design correction (\ref{eq:seqvar2}).}\label{tbl:seqcov}
\end{table}

\section{Discussion}

We presented CIR, a simple, theoretically-motivated modification to IR that promises a substantial improvement in estimation precision when monotonicity violations occur. CIR is widely applicable: the point estimation method described here is adequate for any distribution, not only the Binomial. The only monotone scenarios for which CIR might be less effective than IR are when the true dose-response function $F$ has a staircase-like shape; and that exception only holds for forward estimation. This unusual counter-example helps illustrate that CIR's advantage over IR stems from providing a better fit to the true generic shape of $F$.

CIR's performance advantage over IR is most pronounced in forward estimation of $F$ at interior design points, for which the MSE ratio was $\sim 2$ or more. This might be because everywhere else along a monotonicity violation, in both methods, point estimates pool data from both the left and the right. But at interior design points on the edge of IR's flat stretches, the IR estimate utilizes information only from one side, rendering it half as efficient or worse, on average.

As suggested in the Introduction, when $m$ is sufficiently large for stable estimation of monotone splines, these may be preferable to CIR. However, monotone splines are a very recent and still-evolving field, and therefore even with moderately large $m$, CIR can be useful as a more robust benchmark.

Interval estimation has received a substantial amount of attention in this article. As explained in the Introduction, to date there had been no acceptable small-sample interval estimate for IR; therefore finding one for CIR necessitated developing it \emph{de novo} for isotonic regression in general, at least in the small-sample, conditional-Binomial case. We believe that the combined approach presented here is theoretically backed while having good operating characteristics, and can serve as a basis for further research and improvement. Coverage is generally conservative, except for the sequential dose-finding design we examined. For interval estimation of $F^{-1}$, we strongly recommend the ``local'' interval-inversion approach. The correction (\ref{eq:seqvar2}) for uncertainty in the allocation distribution over $x$ values for sequential designs has only a minor effect, but for specific, analytically-tractable sequential designs it might be improved upon by using a more exact expression.

The R \texttt{cir} package can be downloaded any time from GitHub, under \texttt{`assaforon/cir'}. It offers additional implementation details and user options, such as controlling the behavior on the boundaries.

\subsection*{Funding}

The lead author's work was supported in part by the National Center For Advancing Translational Sciences of the National Institutes of Health under Award Number UL1TR000423. The content is solely the responsibility of the authors and does not necessarily represent the official views of the National Institutes of Health.

\subsection*{Acknowledgments}

The authors thank the Associate Editor and the anonymous reviewers, for constructive comments and questions which have helped improve the manuscript substantially. The lead author would like to thank Marloes Maathuis for her good advice early on in CIR's development, Alexia Iasonos for her helpful response to inquiries regarding the Morris confidence interval, and Michele Shaffer for her support and encouragement to continue methodological work while at Seattle Children's.

\bibliographystyle{elsart-harv}
\bibliography{../dosefinding}

\newpage
\section*{Supplemental Information}

\noindent In the journal version, the following sections were in a separate supplement file.

\subsubsection*{Implementing and Modifying the \citet{Morris88} Ordered-Binomial Interval}

\citet{Morris88} developed a specific iterative solution to the ordered-Binomial interval bound problem that conforms to the conditions of his interval-coverage proof. Assume we have $m$ Binomial summaries at the dose levels $x_1,\ldots ,x_m$, using the article's terminology, i.e., each summary represents $n_my_m$ positive responses out of $n_m$ observations at $x_m$.

For the upper bound, one defines a set of $m$ cumulative distribution functions $G_j(x)$, $j=1,\ldots,m$, starting with the $m$-th function:
$$
G_m\left(y_m\mid n_m,\theta^{UCL}_m\right)=Bin_F\left(n_my_m \mid n_m,\tilde{\theta}^{UCL}_m\right),
$$
\noindent where $Bin_F$ is the Binomial CDF and $\theta^{UCL}_m$ is the upper (forward) confidence bound at $x_m$. We then solve for $\tilde{\theta}^{UCL}_m$ by equating $G_m$ to $\alpha/2$ (with $1-\alpha$ being the specified interval coverage). This produces a UCL equivalent to the Clopper-Pearson bound. For each subsequent dose level indexed $j,j=m-1,\ldots,1$,

\begin{equation*}
\begin{split}
G_j\left(y_j\mid n_j,\theta^{UCL}_j\right)=Bin_F\left(n_jy_j-1 \mid n_j,\tilde{\theta}^{UCL}_j\right) \\
\quad + \quad G_{j+1}\left(\cdot\mid\tilde{\theta}^{UCL}_k\right) Bin_f\left(n_jy_j \mid n_j,\tilde{\theta}^{UCL}_j\right),
\end{split}
\end{equation*}

\noindent where $Bin_f$ is the Binomial probability mass function for exactly $n_jy_j$ positive responses out of $n_j$ observations. This equation defines $G_j$ as a function of $\theta$; the iteration works via the presence of the function $G_{j+1}$. The actual UCL is found as above, by equating $G_j$ to $\alpha/2$. The equations for the LCL are analogous, using CDFs $H_j,j=1,\ldots ,m$, with the iteration proceeding from $H_1$ onwards.

\citet{Morris88}'s formulae had an apparent typo: $G_{j+1}$ in the second equation is written as a function of $\tilde{\theta}^{UCL}_{j+1}$ rather than $\tilde{\theta}^{UCL}_j$ as above. Stated that way, it is already equated to $\alpha/2$, and therefore the iteration cannot proceed. With the formula as above, we were able to reproduce \citet{Morris88} Table~1, which calculated UCLs for the \citet{ReedMuench38} dataset. The correction was verified with Morris (personal communication).

\citet{IasonosOstrovnaya11} used a different method presented by \citet{Morris88}: a generic formula that doesn't use the Binomial probability structure, but rather assumes normal errors. In that method, each bound incorporates a weighted average from dose $j$ and doses to its right (for UCLs) or left (for LCLs). The user has to specify these weights.

The Morris iteration relies upon the Clopper-Pearson bounds due to their direct connection to exact Binomial probabilities, hence ensuring coverage according to \citet{Morris88}'s theorems. However, as mentioned in the article, there exist analytical pointwise Binomial solutions with satisfactory coverage but narrower intervals. Therefore, our code optionally replaces any UCL or LCL produced via the ordered iteration above, with an analytical pointwise bound, in case the latter is tighter. The default alternative is the Wilson bound, but the Agresti-Coull and Jeffrys are also provided in the `cir` package. The latter Agresti-Coull interval is often known as the ``plus four'' interval, because it can be approximated by adding 2 to the numerator and 4 to the denominator of the raw Binomial estimate before calculating the usual asymptotic-theory standard errors.

Finally, bounds are further tightened to enforce monotonicity when applicable. For example, if $\tilde{\theta}^{UCL}_3$ has been further tightened via the Wilson interval, and is now lower than $\tilde{\theta}^{UCL}_2$, then $\tilde{\theta}^{UCL}_2$ will also receive $\tilde{\theta}^{UCL}_3$'s new value.

\subsubsection*{Bias Statistics from Forward Simulations}

\begin{table*}[!ht]
\centering
\begin{tabular}{cccrrrrr}
  \toprule
   \multicolumn{3}{c}{Conditions} & \multicolumn{5}{c}{Pointwise Bias at $x$ Values} \\
 Family & $n$ & Method & $x_2$  & $x_3$ & $x_4$ & ``$x_{2.5}$'' & ``$x_{3.75}$'' \\
  \midrule
\parbox[t]{0mm}{\multirow{6}{*}{\rotatebox[origin=c]{90}{Logistic}}} & \multirow{2}{*}{20} & IR & 0.030 & -0.006 & -0.016 & 0.004 & -0.004 \\
& & CIR &  -0.014 & -0.002 & 0.018 & -0.012 & 0.013  \\
   & \multirow{2}{*}{40} & IR &  0.010 & 0.003 & -0.006 & 0.009 & -0.005 \\
 & & CIR &  -0.009 & -0.001 & 0.006 & -0.000 & -0.000 \\
   & \multirow{2}{*}{80} & IR & 0.001 & -0.001 & -0.004 & 0.003 & -0.006  \\
 & & CIR  & 0.001 & -0.003 & 0.001 & 0.002 & -0.004 \\
  \midrule
\parbox[t]{0mm}{\multirow{6}{*}{\rotatebox[origin=c]{90}{Weibull}}} & \multirow{2}{*}{20} & IR &  0.026 & 0.002 & -0.013 & 0.005 & -0.006 \\
  & & CIR  & -0.034 & -0.018 & 0.021 & -0.025 & 0.006 \\
  & \multirow{2}{*}{40} & IR &  0.006 & 0.004 & -0.002 & 0.003 & -0.002 \\
  & & CIR  & -0.033 & -0.012 & 0.012 & -0.018 & 0.002 \\
  & \multirow{2}{*}{80} & IR &   -0.000 & 0.004 & 0.003 & -0.001 & -0.001 \\
  & & CIR  & -0.020 & -0.009 & 0.004 & -0.013 & -0.003 \\
     \midrule
\parbox[t]{0mm}{\multirow{6}{*}{\rotatebox[origin=c]{90}{``Staircase''}}} & \multirow{2}{*}{20} & IR & 0.011 & 0.000 & 0.004 & -0.025 & 0.021 \\
  & & CIR  & -0.139 & -0.031 & 0.124 & -0.082 & 0.072 \\
  & \multirow{2}{*}{40} & IR &   0.004 & 0.003 & 0.008 & -0.023 & 0.019 \\
  & & CIR  & -0.135 & -0.025 & 0.116 & -0.073 & 0.063 \\
  & \multirow{2}{*}{80} & IR &  0.001 & -0.002 & -0.001 & -0.025 & 0.011 \\
  & & CIR  & -0.136 & -0.032 & 0.106 & -0.073 & 0.056 \\
   \bottomrule
\end{tabular}
\end{table*}

\end{document}